\documentclass[aps, prd,preprint,showpacs,superscriptaddress,groupedaddress,nofootinbib]{revtex4-1}
\usepackage{amsmath,amsfonts}
\usepackage[dvipdfmx]{graphicx}
\usepackage{morefloats}
\graphicspath{{Fig/}}
\begin{document}

\title{Local equilibrium solutions in simple anisotropic cosmological models, as described by relativistic fluid dynamics}
\author{Dmitry Shogin}
\email{dmitry.shogin@uis.no}
\affiliation{Faculty of Science and Technology, University of Stavanger, N-4036 Stavanger, Norway}
\author{Per Amund Amundsen}
\email{per.a.amundsen@uis.no}
\affiliation{Faculty of Science and Technology, University of Stavanger, N-4036 Stavanger, Norway}

\begin{abstract}
We test the physical relevance of the full and the truncated versions of the Israel-Stewart theory of irreversible thermodynamics in a cosmological setting. Using a dynamical systems method, we determine the asymptotic future of plane symmetric Bianchi type~I spacetimes with a viscous mathematical fluid, keeping track of the magnitude of the relative dissipative fluxes, which determines the applicability of the Israel-Stewart theory. We consider the situations where the dissipative mechanisms of shear and bulk viscosity are involved separately and simultaneously. It is demonstrated that the only case in the given model when the fluid asymptotically approaches local thermal equilibrium, and the underlying assumptions of the Israel-Stewart theory are therefore not violated, is that of a dissipative fluid with vanishing bulk viscosity. The truncated Israel-Stewart equations for shear viscosity are found to produce solutions which manifest pathological dynamical features and, in addition, to be strongly sensitive to the choice of initial conditions. Since these features are observed already in the case of an oversimplified mathematical fluid model, we have no reason to assume that the truncation of the Israel-Stewart transport equations will produce relevant results for physically more realistic fluids. The possible role of bulk and shear viscosity in cosmological evolution is also discussed.
\end{abstract}


\maketitle

\section{Introduction}
\label{Sec:Intro}
One of direct consequences of the kinetic theory is that real substances manifest transport properties, which leads to existence of dissipative mechanisms like viscosity and thermal conductivity. Cosmological fluids should not be an exception. Although perfect fluids in local thermal equilibrium (LTE) have been successfully used to model the matter content of the universe, it is dissipation that drives the fluid towards equilibrium: a fluid without appropriate transport properties will never reach this state. Consequently, dissipation must have played an important role in the early universe.
\par 
A brief review of the traditional theories of dissipative thermodynamics can be found in~\cite{Maartens1996}; see references therein for applications to cosmology. At the present moment, the most advanced of these approaches is the second-order Israel-Stewart~(IS) theory~\cite{Israel1976,Israel1979}, often referred to as transient, or causal, thermodynamics. An alternative formalism proposed by Carter~\cite{Carter1991}, see~\cite{Andersson2007} for a review, has been shown to be essentially equivalent to the IS theory~\cite{Priou1991}. 
\par 
The simplified, or "truncated", version of the IS theory is obtained by omitting certain divergence terms in the transport equations of the full theory in a way that does not violate the crucial features of causality and stability. As pointed out by Maartens~\cite{Maartens1996}, the truncated IS theory can be considered as a relativistic generalization of the Navier-Stokes equations for a fluid, in the sense that the values of the transport coefficients are equal to their local equilibrium values. Due to this property, the truncated IS theory is sometimes treated as an independent approach rather than an approximation to the full theory.
\par
Maartens \cite{Maartens1995} has shown that the truncated IS theory can yield solutions which deviate significantly from those of the full theory. Furthermore, Zimdahl \cite{Zimdahl1996a} has demonstated for the case of scalar dissipation that the full and truncated versions are equivalent only in a few special cases. Recently, Shogin et al. \cite{Shogin2015} have discovered that the truncation of the IS transport equations can produce solutions with pathological dynamical features in anisotropic spacetimes. However, despite these results, the truncated theory is still widely used in cosmology, see e.\,g. references in the review by Brevik and Gr{\o}n in~\cite{Travena2013}.
\par 
The standard Friedmann-Robertson-Walker cosmological models have been studied using the truncated \cite{Coley1995,Zimdahl1997a} and the full \cite{Maartens1995, Coley1996} versions of the IS theory. In particular, the processes of inflation and reheating, where dissipative effects play a significant role, have been discussed respectively in~\cite{Maartens1995} and~\cite{Zimdahl1997a}. 
\par
However, the geometry of the spatially homogeneous and isotropic Friedmann-Robertson-Walker model does not allow for any dissipative mechanisms different from bulk viscosity. Therefore, anisotropic backgrounds must be considered when modelling more realistic dissipation. For an early application of causal thermodynamics to anisotropic cosmological models, see Belinskii et al.\,~\cite{Belinskii1979}. Van den Hoogen and Coley \cite{Hoogen1995} have considered Bianchi type~V cosmological models using the truncated IS theory. Heuristical methods have been used to find out that, in contrast to the first-order Eckart theory, the shear viscous stresses play a major role in determining the dynamics of the model; namely, depending on the coefficients of shear viscosity, the spacetime may, or may not, isotropize in the future. Recently, Shogin et al. \cite{Shogin2015} studied Bianchi type IV and V models in the framework both of the full and the truncated versions of the IS theory. The existence of effectively anisotropic solutions, as well as the importance of shear viscosity, was confirmed; in addition, it was demonstrated that even the full IS theory can break down in cosmological applications, as the viscous stresses can drive the fluid far away from LTE. 
\par 
In this paper, we investigate spatially homogeneous plane symmetric Bianchi type~I cosmological models. By this choice we intentionally exclude the effects of spatial curvature to concentrate on determining the limits of applicability of the IS theory, and, in addition, finding the source of the singular behaviour of the solutions of the system in the case when the transport equations are truncated.
\par 
We apply a dynamical systems approach~\cite{Wainwright1997} to describe the future attractors of the system of differential equations governing the dynamics of the cosmological model and perform a stability analysis of the fixed points of this system. Special attention is paid to the late-time behaviour of the relative dissipative fluxes, the magnitude of which represents the deviation of the fluid from LTE, and, by this, determines the applicablity of the IS theory in this particular case.
\par 
The paper is organized as follows. In section~\ref{Sec:Model} we briefly discuss the chosen fluid model. Then, the Hubble-normalized field equations are written down in section~\ref{Sec:Equations}. The future asymptotic states of the cosmological model, as well as the dynamics of the relative dissipative fluxes, are discussed in sections~\ref{Sec:BVSV}--\ref{Sec:SVonly}. Conclusions, together with some possible directions for the further work are presented in section~\ref{Sec:Conslusions}.

\section{The fluid model}
\label{Sec:Model}
We consider the cosmological model to incorporate a non-tilted~\cite{King1973} dissipative fluid, neglecting the effects of thermal conductivity. The energy-momentum tensor of the fluid is then given by
\begin{equation}
T_{\alpha \beta} = (\rho + p + \pi)u_\alpha u_\beta + (p+\pi)g_{\alpha \beta}+\tau_{\alpha \beta},
\end{equation}
where~$\pi$ denotes the bulk viscous pressure, $u_\alpha$ is the four-velocity of the fluid, and~$\tau_{\alpha \beta}$ stands for the shear viscous stress tensor, with~$\tau_{\alpha \beta}u^\beta=\tau_{[\alpha \beta]}=\tau_\alpha^{~\alpha}=0$. 
We are working in the so-called Eckart frame, so the variables $\rho$ and $p$ refer to the local equilibrium values of energy density and pressure of the fluid, and $u_\alpha$ is defined as the {\it particle} four-velocity of the fluid.
\par 
Since our purpose is to perform a qualitative investigation, we consider a {\it mathematical} fluid, for which $\rho$ and~$p$ are connected by a linear barotropic equation of state, widely known as the~$\gamma$-law:
\begin{equation}
p=(\gamma-1)\rho, \qquad 0<\gamma<2.
\end{equation}
Note that the equations of state for {\it physical} radiative fluids are somewhat more complicated, see e.\,g. \cite{Weinberg1971,Straumann1976}, involving temperature and particle number density as independent variables\footnote{A realistic fluid model may, however, lead to simplifications in other aspects of the analysis.}. A simple $\gamma$-fluid with $1<\gamma<4/3$ has been used as a {\it very} naive model of a matter/radiation mixture. 
\par 
The dissipative properties of the fluid are described by the IS theory, the transport equations of the full version\footnote{The transport equation describing heat conduction is omitted.} (in a spatially homogeneous background) being formulated as\footnote{Dots represent derivation with respect to proper time~$t$ of observers comoving with the fluid.}:
\begin{align}
\tau_0\dot{\pi}+\pi &= -3\zeta H-\frac{1}{2}\tau_0\pi \left[3H+\frac{\dot{\tau}_0}{\tau_0}-\frac{\dot{\zeta}}{\zeta}-\frac{\dot{T}}{T} \right],\label{Eq:Model:IS-Full-0-Bulk} \\
\tau_2\dot{\tau}_{ab}+\tau_{ab} &= -2\eta \sigma_{ab}-\frac{1}{2}\tau_2\tau_{ab} \left[3H+\frac{\dot{\tau}_2}{\tau_2}-\frac{\dot{\eta}}{\eta}-\frac{\dot{T}}{T} \right]. \label{Eq:Model:IS-Full-2-Shear}
\end{align}
Here~$T$ is the local equilibrium value for the temperature,~$H$ is the Hubble rate, and~$\sigma_{ab}$ is the geometric rate of shear tensor. In causal thermodynamics, the relaxation times $\tau_0$ and~$\tau_2$ of the dissipative processes are finite and related to the bulk and the shear viscosity coefficients~$\zeta$ and~$\eta$ by

\begin{equation}
\tau_0=\zeta \beta_0, \qquad \tau_2=2\eta \beta_2, \label{Eq:Model:RelaxationTimes}
\end{equation}
where~$\beta_0, \beta_2\geq 0$ are the thermodynamic coefficients for scalar and tensor contributions to the entropy density~\cite{Maartens1996}.

\par 

The transport equations of the truncated IS theory are obtained by dropping the terms in the square brackets on the right-hand sides of~(\ref{Eq:Model:IS-Full-0-Bulk}) and~(\ref{Eq:Model:IS-Full-2-Shear}), which results in:

\begin{align}
\tau_0\dot{\pi}+\pi &= -3\zeta H,\label{Eq:Model:IS-Tr-0-Bulk} \\
\tau_2\dot{\tau}_{ab}+\tau_{ab} &= -2\eta \sigma_{ab}, \label{Eq:Model:IS-Tr-2-Shear}
\end{align}
while the relations (\ref{Eq:Model:RelaxationTimes}) still hold.
\par 
To complete the model, the transient and the transport coefficients must be specified. Following \cite{Hoogen1995} and \cite{Shogin2015}, we assume a barotropic form of the transport coefficients and the relaxation times, the values of the exponents being uniquely determined from dimensional analysis:
\begin{align}
\label{Eq:Model:ViscIndices}
\begin{split}
\zeta & \propto \rho^{1/2}, \qquad \frac{1}{\beta_0} \propto \rho; \\
\eta  & \propto \rho^{1/2}, \qquad \frac{1}{\beta_2} \propto \rho.
\end{split}
\end{align}

Note that for a {\it realistic} fluid these coefficients are more complicated functions of the fluid variables and are derived from radiative thermodynamics \cite{Schweizer1982, Udey1982}. The barotropic approximation may be reasonable in some cases, but here we use it purely for the sake of technical convenience.
\par 
An artefact of these widely adopted models is the need for a separate equation of state for the temperature of the mathematical fluid. For a {\it physical} radiative fluid, no special "temperature model" is needed, as temperature is already incorporated in the equations of state \cite{Schweizer1982,Udey1982}. The two most common choices are \cite{Maartens1996}:
\begin{enumerate}
\item 
The barotropic temperature model, described by 
\begin{equation}
T\propto \rho^{(\gamma-1)/\gamma},
\end{equation}
and used as an approximation for radiation-dominated mixtures ($\gamma$ close to $4/3$);
\item 
The ideal-gas temperature model, with~$p=nT$,
where~$n$ is the (conserved) number density of the fluid obeying
\begin{equation}
\label{Eq:Model:PNumberCons}
\nabla_\alpha(nu^\alpha)=0.
\end{equation}
In Bianchi type~I spacetimes, tensor equation~(\ref{Eq:Model:PNumberCons}) can be rewritten in terms of scalars as
\begin{equation}
\dot{n}+3Hn=0.
\end{equation}
Note that the physical relevance of this approximation is not proven, with a possible exception for ultra-relativistic and highly relativistic ideal gases. 
\end{enumerate}

We use both models in the full version of the IS theory. In the truncated version, however, a particular equation of state for the temperature will not have any influence on the cosmological dynamics, unless a heat-conductive fluid is involved.

\par 
The relative dissipative fluxes caused by the bulk and the shear viscous stresses, respectively, are introduced by:
\begin{equation}
x_\pi=\left \vert \frac{\pi}{p} \right \vert, \qquad x_\tau= \frac{{(\tau_{ab}\tau^{ab})}^{1/2}}{p}.
\end{equation}
The underlying assumption of the IS theory is that the fluid is close to LTE, which implies that these dissipative fluxes are small:
\begin{equation}
\label{Eq:Model:Applicability}
x_\pi,\,x_\tau<<1.
\end{equation}
When condition (\ref{Eq:Model:Applicability}) is violated, the IS theory breaks down, and the resulting solutions cannot be considered physically relevant, although they are mathematically consistent and dynamically stable. We do {\it not} conjecture that the IS theory can be "extrapolated" to produce relevant results when applied to fluids far away from LTE; hence, it is extremely important to keep track of the dissipative fluxes in all the solutions obtained using the IS theory. 

\par 
Finally, it is assumed that the mean interaction time~$t_i$ of the fluid particles is much shorter than the characteristic timescale for macroscopic processes, i.\,e. $t_i<<H^{-1}$, for the hydrodynamic description to be applicable to the
considered matter model.

\section{The system of equations}
\label{Sec:Equations}

We work in the orthornormal frame~\cite{Elst1997} and introduce the scale-independent variables using the common notations of the field, see e.\,g. \cite{Wainwright1997,Ellis2012,Lim2004a}. The normalized energy density~$\Omega$ and the dimensionless geometric shear~$\Sigma_{ab}$ are defined by:
\begin{equation}
\frac{\rho}{3H^2}=\Omega, \qquad \frac{\sigma_{ab}}{H}=\Sigma_{ab}=\text{diag}(-2\Sigma_+,\Sigma_+,\Sigma_+).
\end{equation}
The bulk and the shear viscous stresses, respectively, are normalized by:
\begin{equation}
\frac{\pi}{3H^2}=\Pi, \qquad \frac{\tau_{ab}}{H^2}=\mathcal{T}_{ab}=\text{diag}(-2\mathcal{T}_+,\mathcal{T}_+,\mathcal{T}_+).
\end{equation}
For convenience, we consider Bianchi type~I spacetimes with planar symmetry, for which both~$\Sigma_{ab}$ and~$\mathcal{T}_{ab}$ take the diagonal form. We shall omit the index~"+" in the present paper. Then, the Einstein field equations are\footnote{Primes denote derivation with respect to {\it dimensionless} time $\tilde{t}$, introduced by ${\rm d}\tilde{t}/{\rm d}t=H.$}:
\begin{align}
\Sigma^\prime &= (q-2)\Sigma+\mathcal{T},\\
\Omega^\prime &= (2q+2-3\gamma)\Omega-3\Pi-2\Sigma\mathcal{T},
\end{align}
where
\begin{equation}
q=-\frac{\dot{H}}{H^2}-1=-\frac{H^\prime}{H}-1=2\Sigma^2+\left(\frac{3}{2}\gamma-1 \right)\Omega+\frac{3}{2}\Pi
\end{equation}
is the deceleration parameter. In addition, the Hamiltonian constraint binds~$\Sigma$ and~$\Omega$ algebraically:
\begin{equation}
\label{Eq:Eqs:Hamilton}
1=\Sigma^2+\Omega.
\end{equation}

The system is completed by the dimensionless transport equations, where the bulk and the shear viscosity parameters~$a_0, b_0$ and~$a_2, b_2,$ respectively, are introduced by:
\begin{align}
\begin{split}
\frac{1}{H^2}\cdot\frac{1}{\beta_0} &= a_0 \Omega, \qquad \frac{1}{H}\cdot \frac{1}{\zeta \beta_0}=b_0\sqrt{\Omega},\\
\frac{1}{H^2}\cdot\frac{1}{\beta_2} &= a_2 \Omega, \qquad \frac{1}{H}\cdot \frac{1}{2\eta \beta_2}=b_2\sqrt{\Omega}
\end{split}
\end{align}
and treated as positive constants. The bulk viscosity parameter~$a_0$ is in addition restricted from above by~$a_0<3\gamma(2-\gamma)$, providing that bulk viscous perturbations propagate at finite speeds \cite{Maartens1996}.
\par 
The full transport equations depend on the particular choice of a temperature model. We consider the following options:
\begin{enumerate}
\item 
For the barotropic temperature model, the full transport equations are
\begin{align}
\Pi^\prime &= \left[ -\frac{3}{2}+\frac{1+q}{\gamma}-b_0\sqrt{\Omega}+\frac{2\gamma-1}{2\gamma}\frac{\Omega^\prime}{\Omega}\right]\Pi-a_0\Omega, \label{Eq:System:Baro1} \\
\mathcal{T}^\prime &= \left[ -\frac{3}{2}+\frac{1+q}{\gamma}-b_2\sqrt{\Omega}+\frac{2\gamma-1}{2\gamma}\frac{\Omega^\prime}{\Omega}\right]\mathcal{T}-2a_2\Sigma.\label{Eq:System:Baro2}
\end{align}
\item 
For the ideal-gas temperature model, the full transport equations take the form
\begin{align}
\Pi^\prime &= \left[ -b_0\sqrt{\Omega}+\frac{\Omega^\prime}{\Omega}\right]\Pi-a_0\Omega, \\
\mathcal{T}^\prime &= \left[ -b_2\sqrt{\Omega}+\frac{\Omega^\prime}{\Omega}\right]\mathcal{T}-2a_2\Sigma.
\end{align}
\item 
Finally, the transport equations of the truncated IS theory are:
\begin{align}
\Pi^\prime &= \left[2(q+1)-b_0\sqrt{\Omega}\right]\Pi-a_0\Omega, \\
\mathcal{T}^\prime &= \left[2(q+1)-b_2\sqrt{\Omega}\right]\mathcal{T}-2a_2\Sigma.
\end{align}
\end{enumerate}
Note that the evolution equations for the temperature~$T$ and the particle number density~$n$ decouple from the main system. Hence, the dynamics of these fluid variables is trivial and follows directly from the equation of state and/or particle number conservation. This would not be the case in a model incorporating a realistic radiative fluid.
\par 
Note also that the energy density~$\Omega$ and the shear stress~$\Sigma$ are bounded variables, as follows from the restriction~$\Omega \geq 0$ and the Hamiltonian constraint~(\ref{Eq:Eqs:Hamilton}). However, the state space as a whole is unbounded, since no mathematical restriction is imposed on~$\Pi$ and~$\mathcal{T}$. The dimension of the physical state space is three for the fluids with both bulk and shear viscosity, and two in case when one of these dissipative mechanisms is not taken into account. The state vector then belongs to a subspace of~$S^2\times \mathbb{R}^2$ or~$S^2\times \mathbb{R}$, respectively.

\par 
In each case, we investigate the full system of equations both analytically and numerically. We determine the fixed points of the system and perform an analysis of their {\it local} stability in the future\footnote{All the fixed points of the considered dynamical systems are not listed in this paper; only those which can be locally  stable in the future are discussed.}. This is done by standard analytical methods, see e.\,g. \cite{Hoogen1995, Hervik2005, Shogin2014}, while multiple numerical runs at different sets of model parameters and initial conditions are used to make a conjecture about the {\it global} attractor of the system. In all the numerical simulations, the Hamiltonian constraint is chosen to be initially satisfied.

\par 
The dissipative fluxes, which determine the applicability of the IS theory, are studied by keeping track of the quantities
\begin{equation}
x=\left \vert \frac{\Pi}{\Omega} \right \vert=\vert \gamma-1 \vert x_\pi, \qquad y= \left \vert \frac{\mathcal{T}}{\Omega} \right \vert=3\vert \gamma-1 \vert x_\tau.
\end{equation}

\section{Solutions for bulk/shear viscous fluids}
\label{Sec:BVSV}
In all the three models under consideration, the only locally stable future attractor is described by 
\begin{equation}
\label{Eq:BVSV:Attractor}
[\Sigma,\mathcal{T}, \Omega,\Pi, q]=[0,0,1,\bar{\Pi},\bar{q}],
\end{equation}
with~$\bar{\Pi}<0$. This state describes an izotropizing cosmological model, which is dominated by the bulk viscous fluid at late times. The sign of~$\bar{q}$ is determined by a model-specific relation between~$\gamma$ and the bulk viscosity parameters. Thus, depending on this sign, the spacetime can end up in a state of decelerated, uniform, or accelerated expansion; the latter case represents bulk viscous inflation. Note that anisotropic solutions, which are present in anisotropic cosmological models with non-zero spatial curvature \cite{Hoogen1995,Shogin2015}, are not obtained in Bianchi type~I backgrounds.
\par 
While the numerical runs reveal that all the solutions obtained with the full IS equations tend asymptotically to the state~(\ref{Eq:BVSV:Attractor}), the situation is different when the transport equations are truncated. For a wide range of values of the viscosity parameters, the solutions behave unphysically in the truncated IS theory, running into singularities. We consider this in details in section~\ref{Ssec:BVSV:Trunk}.
\par 
For the non-singular solutions, the relative dissipative flux caused by the shear viscous stresses decays, while that caused by the bulk viscous stresses tends to a negative constant value. In general, this value is not small. Thus, the bulk viscosity can, and generally {\it does}, prevent the fluid from approaching LTE at late times, which is quite typical for similar systems where this dissipative mechanism is involved~\cite{Shogin2015}.

\subsection{The full transport equations, barotropic temperature}
The constants in~(\ref{Eq:BVSV:Attractor}) are given by
\begin{align}
\begin{split}
\label{Eq:BVSV:Constants:Baro}
\bar{\Pi} &= \frac{1}{3}\left[ \gamma b_0 -\sqrt{\gamma^2 b_0^2+6\gamma a_0}\right],\\
\bar{q} &= \frac{1}{2} \left[ 3\gamma-2+\gamma b_0 -\sqrt{\gamma^2 b_0^2+6\gamma a_0}\right].
\end{split}
\end{align}
Calculating the eigenvalues corresponding to the state~(\ref{Eq:BVSV:Attractor}) yields
\begin{align}
\begin{split}
\lambda_1 &= -\frac{\sqrt{\gamma^2 b_0^2+6\gamma a_0}}{\gamma},\\
\lambda_{2,3} &= \frac{3\bar{\Pi}(\gamma+1)+3\gamma(\gamma-2)-2\gamma b_2 \pm \sqrt{\left[3\bar{\Pi}(\gamma-1)+3\gamma(\gamma-2)+2\gamma b_2 \right]^2-32\gamma^2 a_2}}{4},
\end{split}
\end{align}
the rational part of~$\lambda_{2,3}$ being negative. As~$\lambda_1$ is negative, the local stability of the future attractor requires~$\text{Re}(\lambda_{2,3})<0.$ This yields
\begin{equation}
a_2>-\frac{3}{8\gamma}(3\bar{\Pi}-2\gamma b_2)(\bar{\Pi}+\gamma-2),
\end{equation}
which is fulfilled, since the right-hand side is negative.

\subsection{The full transport equations, ideal-gas temperature}
In this case, the future attractor~(\ref{Eq:BVSV:Attractor}) is specified by
\begin{align}
\begin{split}
\label{Eq:BVSV:constants:IG}
\bar{\Pi} &= -\frac{a_0}{b_0},\\
\bar{q} &= \frac{3}{2}\left(\gamma-\frac{2}{3}-\frac{a_0}{b_0}\right),
\end{split}
\end{align}
and the corresponding eigenvalues are given by
\begin{align}
\begin{split}
\lambda_1 &= -b_0,\\
\lambda_{2,3} &=\frac{3(\bar{\Pi}+\gamma-2)-2b_2\pm \sqrt{\left[3(\bar{\Pi}+\gamma-2)+2b_2 \right]^2-32a_2} }{4}.
\end{split}
\end{align}
Again, $\lambda_1$ is negative; the future stability requires~$\text{Re}(\lambda_{2,3})<0$, which leads to
\begin{equation}
a_2>\frac{3}{4}b_2(\bar{\Pi}+\gamma-2).
\end{equation}
The right-hand side is negative, while~$a_2>0$. The future attractor is locally stable for the whole range of values of the model parameters.

\subsection{The truncated transport equations}
Now the constants in~(\ref{Eq:BVSV:Attractor}) are
\label{Ssec:BVSV:Trunk}
\begin{align}
\begin{split}
\label{Eq:BVSV:constants:Trunk}
\bar{\Pi} &= \frac{1}{6}\left[-3\gamma+b_0-\sqrt{(3\gamma-b_0)^2+12a_0}\right],\\
\bar{q} &= -1+\frac{1}{4}\left[3\gamma+b_0-\sqrt{(3\gamma-b_0)^2+12a_0}\right],
\end{split}
\end{align}
the corresponding eigenvalues being
\begin{align}
\begin{split}
\lambda_1 &= -\sqrt{(3\gamma-b_0)^2+12a_0},\\
\lambda_{2,3} &= \frac{3(3\bar{\Pi}+3\gamma-2)-2b_2\pm \sqrt {\left[ 3(\bar{\Pi}+\gamma+2-2b_2) \right]^2-32a_2} }{4}.
\end{split}
\end{align}
The local stability of the future asymptotic state requires~$\text{Re}(\lambda_{2,3})<0;$ the solution of this inequality yields
\begin{align}
\begin{split}
\label{Eq:BVSV:Trunk:Stability}
b_2 &> \frac{3}{2}(3\gamma-2)+9\bar{\Pi},\\
a_2 &> \frac{3}{4} \left[ b_2(\bar{\Pi}+\gamma-2)+6(\bar{\Pi}+\gamma)-3(\bar{\Pi}+\gamma)^2 \right].
\end{split}
\end{align}
The shear viscosity coefficients turn out to play a crucial role in the truncated IS theory. For a wide range of values of these parameters, the requirements~(\ref{Eq:BVSV:Trunk:Stability}) are not fulfilled. In this case, no fixed point of the system is locally stable in the future; numerical simulations show that all the solutions obtained end up in a singularity. The nature of this singularity is that the energy density variable~$\Omega$ crosses the vacuum boundary~($\Omega=0$) and becomes negative. The transport equations are non-defined in this region, and the solutions break down. This is not found to happen in the full IS theory, but singularities of the same type have been reported in more advanced Bianchi models with a dissipative mathematical fluid described by the truncated IS theory~\cite{Shogin2015}.

\begin{figure}[ht!]
\begin{minipage}[ht!]{0.45\linewidth}
\includegraphics[width=0.8\linewidth]{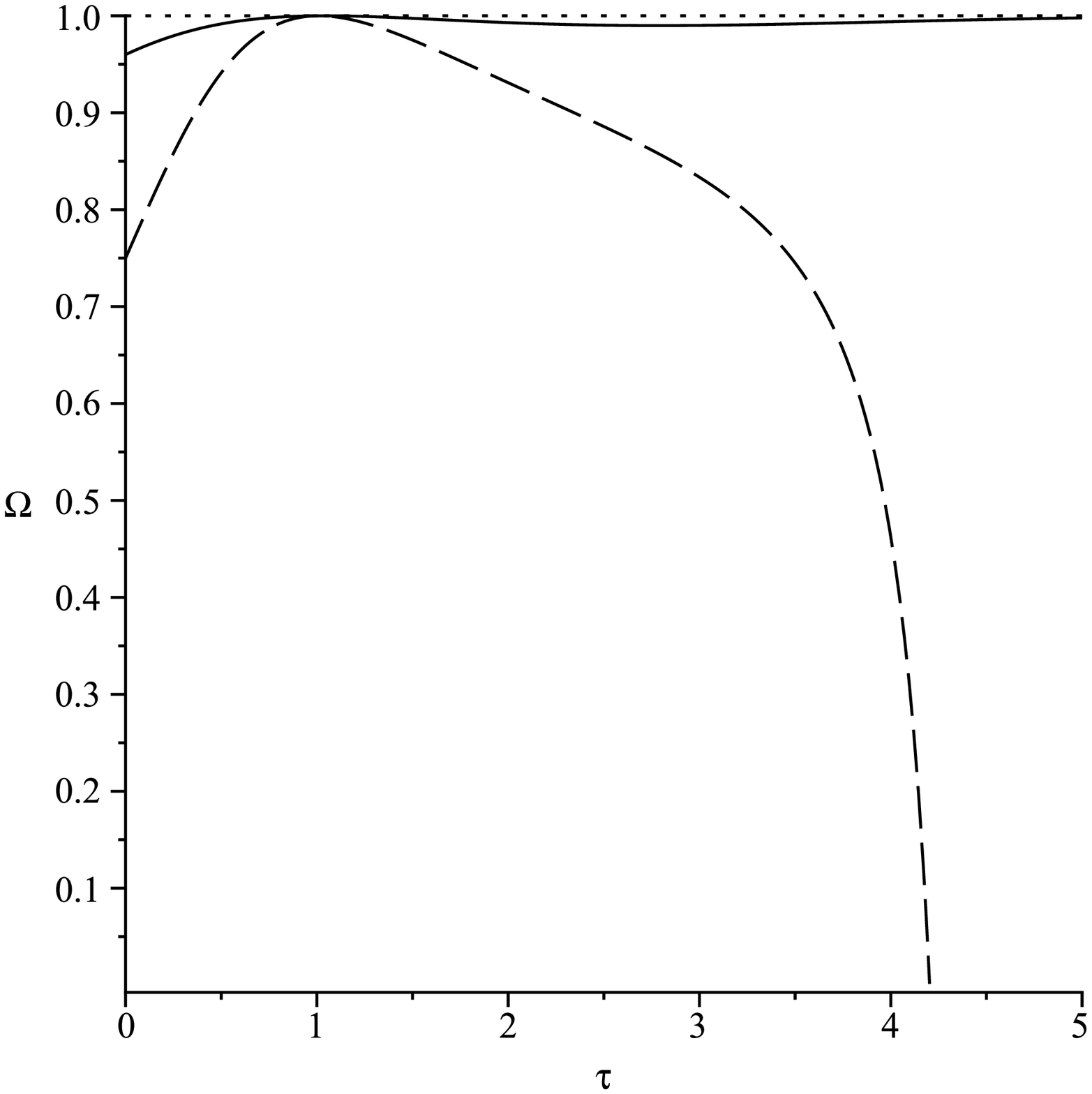}
\end{minipage}
\hfill
\begin{minipage}[ht!]{0.45\linewidth}
\includegraphics[width=0.8\linewidth]{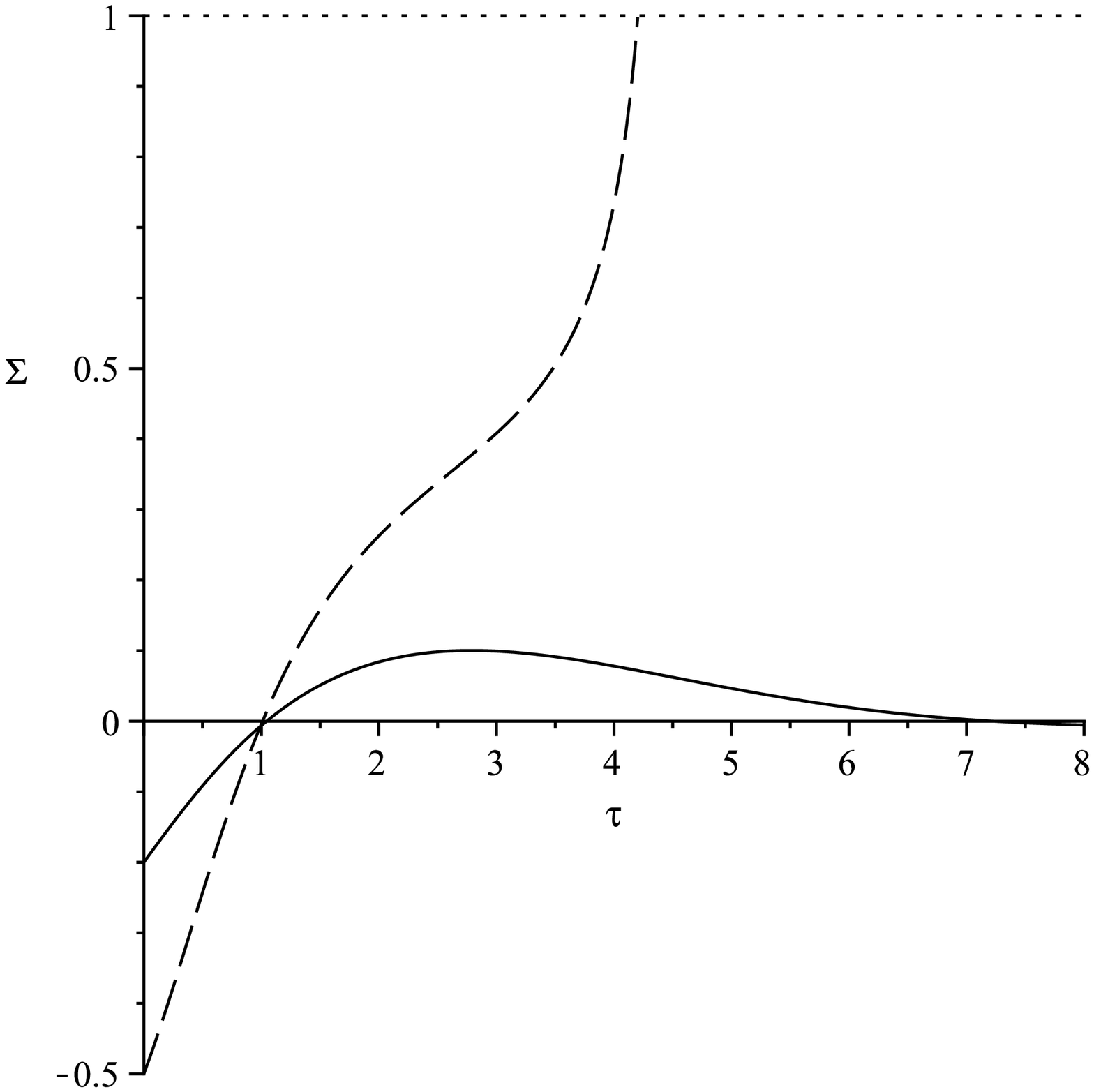}
\end{minipage}
\caption{The dynamics of~$\Omega(\tau)$ (left) and~$\Sigma(\tau)$ (right) in a Bianchi type~I cosmological model with a bulk/shear viscous fluid described by the truncated Israel-Stewart theory. The plots illustrate that the dynamical behaviour of the solutions depends strongly on the initial conditions.\\
At~$\Sigma(0)=-0.2,~\Omega(0)=0.96$ (solid lines), the solution asymptotically approaches the state~(\ref{Eq:BVSV:Attractor}), as might be expected. At~$\Sigma(0)=-0.5,~\Omega(0)=0.75$ (dashed lines), the solution ends up in a singularity, as the energy density variable crosses the vacuum boundary ($\Omega=0$). \\
The model parameters are~$\gamma=6/5,~ a_0=0.1,~b_0=5.0,~a_2=0.5,~b_2=3.0$, so the inequalities~(\ref{Eq:BVSV:Trunk:Stability}) are satisfied. }
\label{Fig:PT}
\end{figure}

\par 
Moreover, the system with the truncated transport equations is found to be sensitive to alterations of the initial conditions. That is, it is possible to obtain singular solutions even if the viscosity parameters {\it do} satisfy the inequalities~(\ref{Eq:BVSV:Trunk:Stability}). An example demonstrating this feature is shown in Figure~\ref{Fig:PT}.

\par 
For the cosmological model considered, such singular solutions do not exist in the full IS theory. Thus, it is the truncation of the IS equations which allows the energy density to manifest unphysical behaviour.

\section{Solutions for bulk viscous fluids with vanishing shear viscosity}
\label{Sec:BVonly}
In the absence of shear viscosity the future attractor retains the form~(\ref{Eq:BVSV:Attractor}), assumed~$\mathcal{T}\equiv 0$. The expressions for $\bar{\Pi}$ and~$\bar{q}$, obtained in section~\ref{Sec:BVSV}, also hold in this case. However, the instabilities of the truncated IS theory described in section~\ref{Ssec:BVSV:Trunk} are removed.
\par 
Although the future attractor is essentially the same, the dynamics of the approach to it is changed, as the corresponding eigenvalues are different from those calculated in section~\ref{Sec:BVSV}.
\par 
It turns out that the shear viscous stresses play only a minor role in the full IS theory, provided the bulk viscosity is non-zero; namely, they affect neither the future asymptotic state nor its stability. On the other hand, the role of shear viscosity in the truncated version of the IS theory is crucial, which is consistent with the results of \cite{Hoogen1995}.
\par 
The relative dissipative flux caused by the bulk viscous stresses tends to a finite constant at late times: in general, the fluid does not asymptotically approach LTE. 

\subsection{The full transport equations, barotropic temperature}
The future asymptotic values of~$\Pi$ and $q$ are given by~(\ref{Eq:BVSV:Constants:Baro}); the corresponding eigenvalues are
\begin{equation}
\lambda_1 = -\frac{\sqrt{\gamma^2b_0^2+6\gamma a_0}}{\gamma}, \qquad \lambda_2 = \frac{3}{2}\left[ \bar{\Pi}+(\gamma-2) \right].
\end{equation}
Since~$\bar{\Pi}<0$ and~$\gamma<2$, both eigenvalues are negative reals, and the asymptotic state is locally stable in the future.

\subsection{The full transport equations, ideal-gas temperature}
The values of~$\bar{\Pi}$ and~$\bar{q}$ are provided by~(\ref{Eq:BVSV:constants:IG}), the eigenvalues being
\begin{equation}
\lambda_1 = -b_0, \qquad \lambda_2 = \frac{3}{2}\left[ \bar{\Pi}+(\gamma-2) \right].
\end{equation}
Again, the eigenvalues are negative real numbers, and the future attractor is locally stable.

\subsection{The truncated transport equations}
The constants in~(\ref{Eq:BVSV:Attractor}) are given by~(\ref{Eq:BVSV:constants:Trunk}). The corresponding eigenvalues are
\begin{equation}
\lambda_1 = -\sqrt{(3\gamma-b_0)^2+12a_0}, \qquad \lambda_2 = \frac{3}{2}\left[ \bar{\Pi}+(\gamma-2) \right].
\end{equation}
In contrast to the case of section~\ref{Ssec:BVSV:Trunk}, the eigenvalues are now real and negative. The instability of the future attractor is removed. The singular solutions, which appear in the case of nonvanishing shear viscosity, cannot be obtained in this case. 
\par 
Hence, the pathological dynamical features of solutions in the truncated IS theory originate from the evolution of the shear viscous stresses.

\section{Solutions for shear viscous fluids with vanishing bulk viscosity}
\label{Sec:SVonly}
The only future stable stationary point of the system is given by
\begin{equation}
\label{Eq:SV:Attractor}
[\Sigma, \mathcal{T}, \Omega, q]=[0,0,1,\bar{q}],
\end{equation}
with
\begin{equation}
\bar{q}=\frac{1}{2}(3\gamma-2).
\end{equation}
The asymptotic value of the deceleration parameter depends on~$\gamma$ only and is positive for typical $\gamma$-fluids. Hence, the shear viscosity does not contribute to accelerating the expansion of the universe; however, spatial anisotropy is eliminated at late times.
\par 
For the full IS transport equations, the eigenvalues are either negative reals or complex conjugates with a negative real part; this provides the local stability of the future attractor. For the truncated transport equations, the situation is different: the negativity of~$\text{Re}(\lambda_{1,2})$ can be violated, which makes the state~(\ref{Eq:SV:Attractor}) unstable in the future. The corresponding solutions are found to end up in a singularity, which is discussed in section~\ref{Ssec:SV:Trunk} below.
\par 
For all the non-singular solutions, the relative dissipative fluxes caused by the shear viscous stresses decay in the future, and the fluid approaches LTE at late times. This is the only case in the considered cosmological model when the underlying assumption~(\ref{Eq:Model:Applicability}) of the IS theory is not violated and the solutions of the system can be considered fully reasonable.

\subsection{The full transport equations}
The eigenvalues corresponding to the state~(\ref{Eq:SV:Attractor}) are the same for both temperature models:
\begin{equation}
\lambda_{1,2}=\frac{(3\gamma-2b_2-6) \pm \sqrt{(3\gamma+2b_2-6)^2-32a_2}}{4},
\end{equation}
where the rational part is negative. The local stability requirement~$\text{Re}(\lambda_{1,2})<0$ yields
\begin{equation}
a_2>\frac{3}{4}b_2(\gamma-2),
\end{equation}
 which is fulfilled automatically by~$a_2>0$ and~$\gamma<2$.

\subsection{The truncated transport equations}
\label{Ssec:SV:Trunk}
For the truncated transport equations, the eigenvalues are given by
\begin{equation}
\lambda_{1,2}=\frac{ (9\gamma-2b_2-6)\pm \sqrt{(3\gamma-2b_2+6)^2-32a_2}}{4}.
\end{equation}
An algebraic analysis of the local stability requirement~$\text{Re}(\lambda_{1,2})<0$ results in the following system of inequalities:

\begin{align}
\begin{split}
\label{Eq:SV:Restrict}
b_2 &> \frac{3}{2}(3\gamma-2),\\
a_2 &> \frac{3}{4}(\gamma-2)(b_2-3\gamma).
\end{split}
\end{align}
If the values of~$a_2$ and~$b_2$ do not satisfy these conditions, the corresponding solutions behave unphysically and end up in a singularity of the same nature as described in section~\ref{Ssec:BVSV:Trunk}. Moreover, similarly to the case with nonvanishing bulk and shear viscosity, the stability of the future attractor depends crucially on the initial conditions; it is possible to obtain singular solutions, even if the inequalities~(\ref{Eq:SV:Restrict}) are satisfied. As might be expected, this is not a property of the system with the full transport equations.
\par 
This confirms our assumption that it is the evolution equations for the shear viscous stresses that lead to unacceptable properties of the solutions obtained using the truncated IS theory.

\section{Conclusions}
\label{Sec:Conslusions}

We have used the dynamical systems approach to investigate the future attractors and their stability conditions for viscous mathematical fluids in Bianchi type~I spacetimes. We have studied the properties of cosmological solutions obtained with the full and the truncated versions of the IS~theory, having used two simple temperature models in the full version. Also, we have determined the asymptotic future of the relative dissipative fluxes to find out when the near-equilibrium conditions are violated and the IS~theory breaks down.
\par 
All the solutions obtained using the full IS~theory are found to be non-singular in future and stable under alterations of the initial conditions. On the contrary, the truncated IS transport equations, if applied to a fluid with nonvanishing shear viscosity, allow the energy density to cross the vacuum boundary, which leads to unphysical, singular behaviour of the solutions already in the simplest anisotropic spacetimes.
\par 
Another pathological feature of the truncated IS equations discovered in the present work is the extreme sensitivity of the resulting system to the choice of the initial conditions. When the truncated IS theory is applied to a $\gamma$-fluid with nonvanishing shear viscosity, the solutions can run into a singularity even if there exists a stable future asymptotic state.
\par 
The solutions of the full IS theory describe an izotropizing universe dominated by the dissipative fluid in the asymptotic future. The shear viscous stresses, if present, decay at late times, while the bulk viscous stress, if present, freezes into a negative constant value. In these solutions, the bulk viscosity eliminates spatial anisotropy and, in addition, can accelerate the expansion of the universe. The shear viscosity does not contribute to accelerating the expansion, but effectively eliminates the Bianchi type~I anisotropy even when bulk viscosity is zero. Anisotropic solutions, which exist in more complicated spacetimes~\cite{Hoogen1995,Shogin2015}, cannot be obtained in Bianchi type~I cosmological models.
\par
The full IS theory provides a completely reasonable description {\it only} for shear viscous mathematical fluids with vanishing bulk viscosity: the relative dissipative flux caused by the shear viscous stresses decays exponentially in the future, and the fluid is driven towards LTE. 
\par
The full IS theory yields stable solutions for bulk viscous fluids. However, these solutions are not fully consistent with the underlying assumptions of the IS theory, since generally the fluid neither is close to LTE during its evolution nor approaches thermal equilibrium in the asymptotic future. The near-equilibrium conditions can be violated by the bulk viscosity: the corresponding relative dissipative flux is asymptotically a negative constant, which is not small in general. This departure from LTE is, of course, a physical result, which may depend on the fluid model adopted, and per se not a failure of the IS theory.
\par
Large deviations from LTE caused by the shear viscous stresses, which have been discovered in more general cosmological models~\cite{Shogin2015}, are not present in Bianchi type~I spacetimes.
\par  
The future asymptotic states of the Bianchi type~I spacetimes are of similar form for the two temperature models considered in the full IS~theory. Still, the solutions can behave differently in the future; for example, these models under the same initial conditions can result in opposite signs of the asymptotic value of the deceleration parameter~$q$. Also, the dynamical character of the solutions (monotoneous or oscillatory behaviour) can differ between them.
\par 
The results obtained suggest that in solving real physical problems, the full IS theory should be preferred over its truncated version. For a mathematical fluid with barotropic transport coefficients, the bulk viscosity creates finite, non-decaying relative dissipative fluxes, which lead to a possible breakdown of the full IS theory; hence, consistent non-linear thermodynamical theories, which can describe fluids substantially away from LTE, are potentially important. However, for realistic radiative fluids with transient and transport coefficients derived from the kinetic theory, the dynamics of the variables may be essentially different. We leave considering physical fluids in anisotropic cosmological backgrounds for the further investigations.
\bibliography{Bib/New}
\end{document}